Research Article

Aditya Tripathi, Sergey Kruk*, Yunfei Shang, Jiajia Zhou, Ivan Kravchenko, Dayong Jin, and Yuri Kivshar

# Topological nanophotonics for photoluminescence control

**Abstract:** Rare-earth doped nanocrystals are emerging light sources used for many applications in nanotechnology enabled by human ability to control their various optical properties with chemistry and material science. However, one important optical problem – polarisation of photoluminescence – remains largely out of control by chemistry methods. Control over photoluminescence polarisation can be gained via coupling of emitters to resonant nanostructures such as optical antennas and metasurfaces. However, the resulting polarization is typically sensitive to position disorder of emitters, which is difficult to mitigate. Recently, new classes of disorder-immune optical systems have been explored within the framework of topological photonics. Here we explore disorder-robust topological arrays of Mie-resonant nanoparticles for polarisation control of photoluminescence of nanocrystals. We demonstrate polarized emission from rare-earth-doped nanocrystals governed by photonic topological edge states supported by zigzag arrays of dielectric resonators. We verify the topological origin of polarised photoluminescence by comparing emission from nanoparticles coupled to topologically trivial and nontrivial arrays of nanoresonators.

**Aditya Tripathi, Sergey Kruk, Yuri Kivshar,** Nonlinear Physics Center, Research School of Physics, Australian National University, Canberra ACT 2601, Australia
**Aditya Tripathi,** Department of Physics, Indian Institute of Technology Delhi, New Delhi 110016
**\*Corresponding author: Sergey Kruk,** Department of Physics, University of Paderborn, Paderborn D-33098, Germany
**Yunfei Shang, Jiajia Zhou, Dayong Jin,** Institute for Biomedical Materials and Devices (IBMD), University of Technology Sydney, Sydney NSW 2007, Australia
**Ivan Kravchenko,** Center for Nanophase Materials Sciences, Oak Ridge National Laboratory, Oak Ridge, TN 37831, USA

# 1 Introduction

Topological phases of light provide unique opportunities to create photonic systems immune to scattering losses and disorder [1]. Motivated by on-chip applications, there have been efforts to bring topological photonics to the nanoscale. Nanostructures made of high-index dielectric materials with judiciously designed subwavelength resonant elements supporting both electric and magnetic Mie resonances [2] show a special promise for implementations of the topological order for light.

Topological structures have recently been studied as a powerful tools for harnessing light emission in various systems with topological orders including lasers [3, 4], quantum light sources [5], and nonlinear frequency converters [6]. In addition, topological photonics holds promise for the development of novel types of light emitters, as it provides a systematic way to control the number and degree of localization of spectrally isolated topologically-robust edge and corner states.

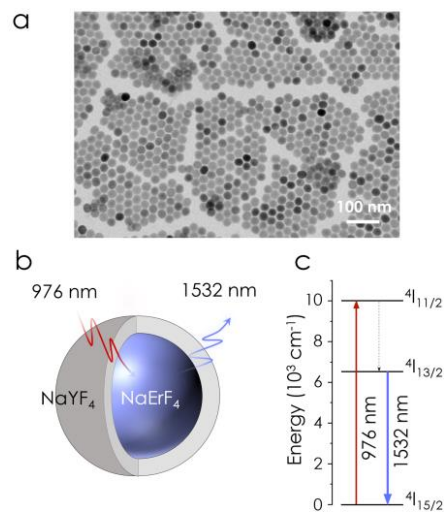

**Fig. 1: Structure and properties of $Er^{3+}$-doped core-shell nanoparticles.** (a) Transmission electron microscope image of nanoparticles. (b) Schematic of the core-shell structure of the nanoparticle. (c) Relevant energy levels of $Er^{3+}$.

Here we uncover *another class of nontrivial effects in harnessing light emission with topology*. We employ topological properties of nanoscale photonic systems to control polarisation of photoluminescence of rare-earth-doped nanocrystals.

Rare-earth-doped nanocrystals are emerging light sources used for many applications of nanotechnology like bio-imaging, sensing, therapy, display, data storage, and photonics devices [7, 8]. The applicability of nanocrystals in the above aspects is determined by human ability to control their performance in diverse optical dimensions including emission color, spectrum, lifetime, intensity, and polarization.

To date, except for polarization, all other optical parameters of the rare-earth-doped nanocrystals have been extensively engineered by chemistry and material science methods. [9–12]. Though polarized emission of rare-earth nanocrystals has been observed [13–16] and demonstrated for flow shear tomography [17], the polarization anisotropy was only detectable through spectroscopic measurements from single or aligned crystals, and most commonly, rod-type nanocrystals. This is because the polarization anisotropy is assigned to each splitting transition of rare-earth ions, and it is determined by the site symmetry in a crystal host. The irregular orientation of the crystalline axis will lead to a neutralization effect. The whole-band based imaging will also neutralize the polarization information, as narrow-peaks from 4f-4f rare-earth transitions may have different dipole orientations. A control of the polarization is even more challenging, as a given nanocrystal always has a fixed crystal phase and site symmetry.

To gain a control over polarization of emission, the emitters can be coupled to resonant nanostructures. Coupling of emitters to individual nanoantennas [18–24] and metasurfaces [25–27] has already been explored for control over the polarization of light [28, 29]. However, the resulting polarization of emission depends typically on the specific positioning of emitters. Precise positioning of emitters on nanostructures is challenging, and typically it requires demanding approaches such as atomic force microscopy [30]. In this regard, polarization control that relies on topologically nontrivial optical modes in nanostructures becomes attractive as topology introduces robustness against disorder in positioning. The topological protection offers to mitigate the dependence of the system on the exact position and orientation of emitters.

In this paper, we employ topological nanophotonics to control the emission polarization of the rare-earth doped nanocrystals. Specifically, we employ zigzag arrays of dielectric nanoresonators hosting topologically nontrivial optical modes that are robust against perturbations of the system [6]. We couple them with $Er^{3+}$-doped nanocrystals and observe enhanced polarized photoluminescence (PL).

We reveal that topological edge states can control polarization of emission in an unusual way. Specifically, in the vicinity of topological edge states the emission becomes linearly polarized reproducing the polarization of topological edge modes. For the arrays with odd number of nanoresonators, the PL emission from two edges is orthogonally polarized, and for the arrays with even number of nanoresonators the PL emission becomes co-polarized, in accord with the polarization of the topological states. To test the topological origin of the polarized emission, we study topologically trivial arrays and observe in contrast completely depolarized emission.

## 2 Results and Discussion

We use β-$NaErF_4$@$NaYF_4$ core-shell nanocrystals with average diameter of 25 nm, as shown in the TEM image in Fig.1a and in the schematics in Fig.1b. The nanocrystals are synthesized by employing the layer epitaxial growth method [31]. When pumped with 976 nm wavelength laser, the nanocrystals produce photoluminescence at around 1532 nm wavelength, which corresponds to the $Er^{3+}$ transition: $^4I_{13/2}$ ---→ $^4I_{15/2}$, as shown with the relevant energy levels in Fig.1c. The geometry of the zigzag nanostructures is chosen such that the wavelength of the topological edge states matches the wavelength of the nanoparticles' photoluminescence. In the topological zigzag array, each disk hosts Mie-resonant modes [2] [see Fig.2b for details about multipolar expansion and Figs.2c for near-field distributions of the scattered electric field]. The electric dipole mode of the disk peaks at around the wavelength of the $Er^{3+}$ electric dipole transition $^4I_{13/2}$ ---→ $^4I_{15/2}$, thus allowing for coupling between the emitter and the nanostructure.

When such nanoresonators are arranged into zigzag arrays, the geometry of the array introduces altered strong and weak near-neighbor coupling which can be described with a polarization-enriched generalized Su–Schrieffer–Heeger (SSH)-type model [32–34] with a gauge-independent $\mathcal{Z}(2)$ topological invariant. The formation of topologically nontrivial light localization can be described with the analytical coupled dipoles approximation [32]. The Hamiltonian of the system can be written as:

$$\mathcal{H} = \sum_{j,v} \hbar\omega_0 a_{jv}^\dagger a_{jv} + \sum_{\langle j,j'\rangle, v,v'} a_{jv}^\dagger V_{v,v'}^{\langle j,j'\rangle} a_{jv}$$

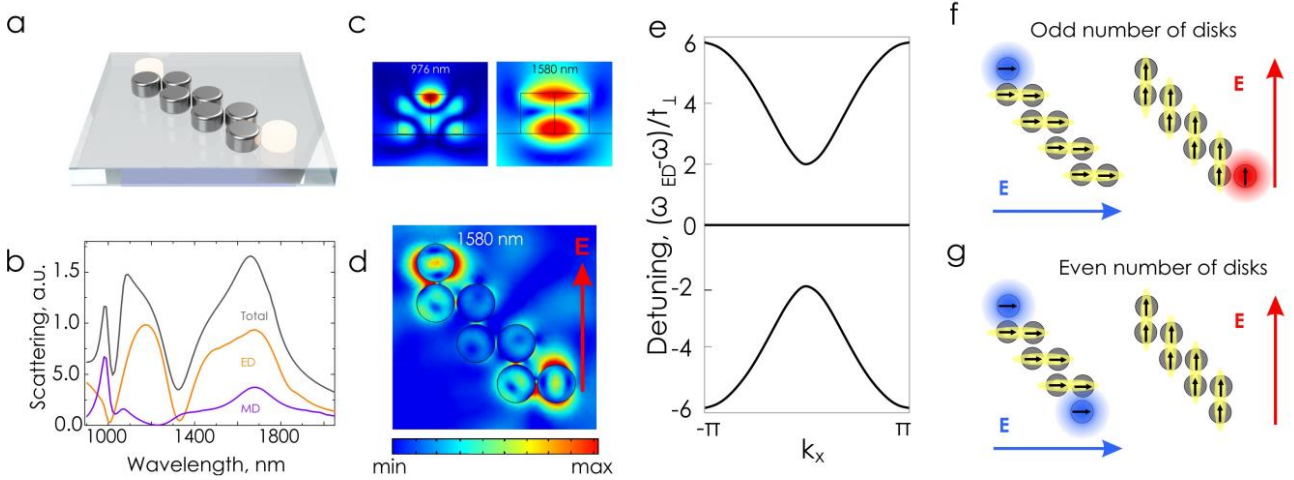

**Fig. 2: Topological zigzag arrays of dielectric nanoresonators.** (a) Concept image of the zigzag array hosting topological edge states. (b) Multipolar decomposition of a constituent disk nanoresonator. (c) Near-field distributions of a single disk for plane wave excitation at around the pump and the photoluminescence wavelengths. (d) Full-wave simulation of edge localizations in the zig-zag array illuminated with a diagonal polarisation. (e) Dispersion curve for Bloch modes in periodic zigzag chain (f ,g) Schematics of the edge states formation in arrays with (f) even and (g) odd number of nanoresonators. Blue and red colours are used to distinguish between two principal polarisation of excitation: horizontal and vertical. Yellow joints visualise stronger dipole-dipole coupling between the modes of the near-neighbour disk for the cases of horizontal polarisation excitation (left) and vertical polarisation excitation (right). Topological states (left side, blue color – horisontally polarised; right side, red color – vertically polarised) form at the edge disks that are weakly coupled to their near-neighbours.

where $\omega_0$ denotes the resonance frequency, the indices $j$ and $j'$ label the nanoparticles, $\langle j,j' \rangle$ are the first nearest neighbors in the array, and $a_{j\nu}$ is the annihilation operator for the multipolar eigenmodes with the polarization $\nu$ at the $j^{th}$ nanoparticle. We consider the electric dipole modes polarized in the plane of the sample $(x,y)$ in either $x$- or $y$ directions. The coupling matrices $V^{\langle j,j' \rangle}$ are then written as:

$$V^{\langle j,j' \rangle} = t_{\parallel} e_{\parallel}^{\langle j,j' \rangle} \otimes e_{\parallel}^{\langle j,j' \rangle} + t_{\perp} e_{\perp}^{\langle j,j' \rangle} \otimes e_{\perp}^{\langle j,j' \rangle}$$

and $e_{\perp}^{\langle j,j' \rangle}$ are the in-plane unit vectors parallel and perpendicular to the vector linking the near neighbor particles and $t_{\parallel}$ and $t_{\perp}$ are the coupling constants for the modes, co-polarized and cross-polarized with respect to the link vector and the $\otimes$ sign stands for the direct product. For short-range dipole–dipole interaction, the ratio of the coupling constants can be estimated as $t_{\parallel}/t_{\perp}=-2$ as shown in Ref. [32]. Figure 2d shows a full-wave simulation of the edge localizations in the zig-zag array.

Figure 2e further visualises the energy spectrum of the zigzag array of coupled resonators. The spectrum features a band-gap with two degenerate edge states associated with the near-field distribution shown in Fig.2d. Non-zero longitudinal components of $k$-vector were accounted in the coupling constants $t \rightarrow t/2 - e^{ik_x l}$ [6]. The edge state energy corresponds to the energy of a stand-alone nanoparticle $\hbar\omega_0$. The resulting topological invariant (winding number) equals 1 within the band gap, which corresponds to the emergence of one polarisation-dependent state at every edge [32]. Figs.2f,g visualize resonant coupling of the electric dipole resonant modes of the neighboring nanodisks. Depending on the mutual orientation, the neighboring dipoles couple either strongly (visualized in Figs. 2 f,g with yellow joints) or weakly. In every case, the weakly coupled edge disk hosts an edge mode.

The resulting edge modes are protected by topology against perturbations of the system such as disorder [6]. The modes are linearly polarized. For arrays with odd number of disks the modes at the opposite ends are orthogonally polarized (see Fig. 2f), and for the even number of disks, the modes are co-polarized (see Fig. 2g).

We proceed to experiments and fabricate the zigzag arrays of nanoparticles from amorphous silicon on a fused silica substrate (500 μm thick). First, a 300 nm amorphous silicon layer was deposited onto the substrate by low pressure chemical vapor deposition. Subsequently, a thin layer of an electron-resist PMMA A4 950 was spin-coated onto the sample, followed by electron beam lithography and development. A thin Cr film was evaporated onto the sample, followed by a lift-of process to generate a hard mask. Reactive-ion etching was used to transfer the Cr mask pattern into the silicon film. The residual Cr mask was removed via wet etching. This resulted in zigzag arrays of disks 510 nm in diameter with 30 nm spacing between the near-neighbors. Finally, Er$^{3+}$-doped nanocrystals were spin-coated on top of the arrays of dielectric nanoresonators.

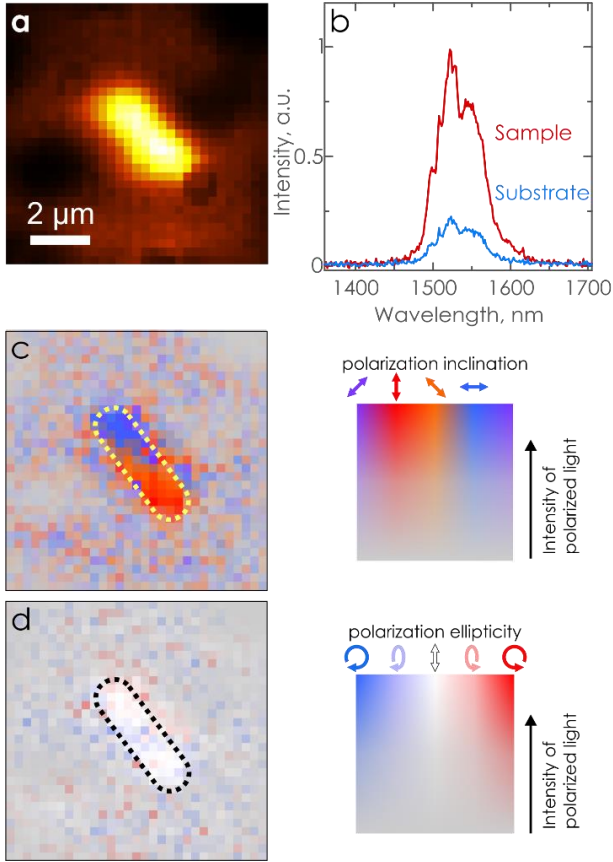

**Fig. 3: Topology-controlled photoluminescence of Er$^{3+}$ coreshell nanoparticles.** (a) Camera image of the emission enhancement from the Er$^{3+}$ nanoparticles in the vicinity of the zigzag array. (b) Photoluminescence spectra of the Er$^{3+}$ nanoparticles on top of a zigzag array vs on top of a bare substrate. (c, d) Spatially-resolved polarization states of photoluminescence showing (c) polarization inclination angles and (d) ellipticity of photoluminescence.

In our optical experiments, we pumped the samples with 976 nm CW diode laser. We narrow down the collimated laser output beam with a telescope made of f=125 mm lens and an objective lens Mitutoyo Plan Apo NIR HR (X100, 0.7NA). We studied spatial distribution of the PL signal in reflection on a camera Xenics Bobcat-320. The resulting measurement is shown in Fig. 3a. We next measured PL spectra in reflection with a spectrometer NIR-Quest from Ocean Optics. Figure 3b shows the comparison of a spectrum measured from nanoparticles on a single array with 9 nanodisks compared to a spectrum of nanoparticles on a bare glass substrate next to the array. **The array provides approximately five-fold enhancement** of PL signal.

Next, we experimentally measured spatially resolved polarization states of PL shown in Figs. 3c,d. For this we introduced a quarter-wave plate and a polarizer; and performed a Stokes-vector polarimetry. Figure 3c visualizes the polarization inclination angle of PL around the zigzag array. Here, saturation visualizes the intensity of polarized emission [gray– no polarized emission] and the colors visualize the inclination angle of the electric field. Red color stands for vertical polarization, and blue for horizontal polarization. Figure 3d visualizes ellipticity of emission where similarly saturation reflects the intensity of polarized emission. We notice that ellipticity of polarization is close to zero in all our experiments. We observe that PL from rare-earth doped nanoparticles becomes polarized in the vicinity of the topological edge states with the polarization state of PL reproducing the polarization of the topological edge mode.

Finally, we compare PL from nanoparticles around the 9-nanodisk zigzag array with 15-nanodisk and 14-nanodisk zigzag arrays as well as with topologically trivial straight array (see Fig. 4). The 15-nanodisk zigzag performs qualitatively similar to the 9-nanodisk array as it has same parity of disks (odd). A longer array exhibits sharper localisation of the PL as it is expected to provide higher intensity contrast with the bulk [6]. In the 15 disk-long chain PL away from the edge becomes depolarised as the topological modes responsible for polarisation control are pinned to the edges. The 14-nanodisk zigzag array in contrast demonstrates co-polarized PL in the vicinity of the two edge states in agreement with its parity (even). The bulk of the 15 and 14-nanodisk arrays produces depolarised photoluminescence due to lack of polarisation sensitivity of zigzag bulk modes. No polarized PL is observed for the topologically trivial array of nanoresonators.

## 3 Conclusion and Outlook

We have studied emission of Er$^{3+}$-doped nanocrystals deposited on top of topologically nontrivial zigzag arrays of Mie-resonant dielectric nanoparticles. We have observed a novel effect in topological photonics: enhanced and polarised photoluminescence emission from the doped nanocrystals located near the topological edge states. We have verified the topological origin of this controlled emission by conducting additional experiments with trivial arrays of the same dielectric nanoresonators, for which we have observed completely depolarized emission.

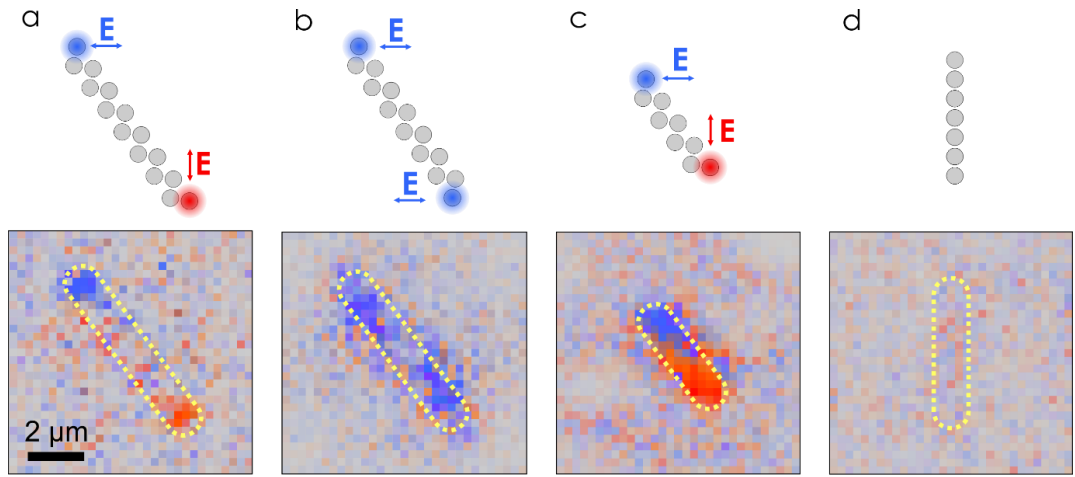

**Fig. 4: Polarization control of photoluminescence with topological edge states.** Spatially resolved polarization states of photoluminescence for (a) 15-nanodisk zigzag array, (b) 14-nanodisk zigzag array, (c) 9-nanodisk zigzag array, and (d) topologically trivial straight array of resonant nanodisks. The color-map is analogous to that used in Fig. 3c.

As the next step in this direction, we expect an efficient control of photon up-conversion being crucial for many applications ranging from bio-imaging to photovoltaics. By adjusting nanostructure parameters, we expect to achieve a systematical shift of the spectral position of resonances over several hundreds of nanometers, and also observe multi-fold enhancement of photon up-conversion efficiency in nanocrystals which is useful for low-threshold photon up-conversion in future solar energy applications.

We expect that our results may open a new direction in the study of topology-enable emission properties of topological edge states in many photonic systems. Indeed, active topological photonics can provide a fertile platform for not only studying interesting fundamental problems involving nontrivial topological phases, but in addition it may become a route towards the development of novel designs for disorder-immune active photonic device applications, such as robust high-speed routing and switching, nanoscale lasers, and quantum light sources.

**Acknowledgment:** A part of this research was conducted at the Center for Nanophase Materials Sciences, which is a DOE Office of Science User Facility. A.T. and S.K. thank Barry Luther-Davies for useful discussions and experimental support. S.K. thanks Alexander Poddubny for useful discussions and also acknowledges a financial support from the Alexander von Humboldt Foundation.

**Funding:** Australian Research Council (grants DP200101168 and DE180100669), Strategic Fund of the Australian National University, and China Scholarship Council (grant 201706120322).